\documentclass[aps,prl,showpacs,english,10pt,a4,nofootinbib,notitlepage,twocolumn,superscriptaddress]{revtex4-1}

\usepackage[utf8]{inputenc}
\usepackage[T1]{fontenc}
\usepackage[english]{babel}

\usepackage{times}
\usepackage{dsfont}

\usepackage{amssymb,amsmath,amsfonts,amsthm}
\usepackage{mathtools}
\usepackage{verbatim}
\usepackage{enumerate}
\usepackage{graphicx}
\graphicspath{{figs/}}
\usepackage{hyperref}
\usepackage{bbm}
\usepackage{booktabs}

\usepackage[caption=false]{subfig}

\usepackage{color}
\definecolor{dred}{rgb}{.8,0.2,.2}
\definecolor{ddred}{rgb}{.8,0.5,.5}
\definecolor{dblue}{rgb}{.2,0.2,.8}
\definecolor{dgreen}{rgb}{.2,0.5,.2}

\newcommand{\bra}[1]{\mbox{$\langle #1|$}}
\newcommand{\ket}[1]{\ensuremath{|#1\rangle}}

\newcommand{\be}{\begin{equation}}
\newcommand{\ee}{\end{equation}}

\newcommand{\bea}{\begin{eqnarray}}
\newcommand{\eea}{\end{eqnarray}}

\begin{document}

\title{Measurement of linear response functions in Nuclear Magnetic Resonance}
\date{\today}
\author{Tao Xin}
\affiliation{State Key Laboratory of Low-dimensional Quantum Physics and Department of Physics, Tsinghua University, Beijing 100084, China}
\affiliation{Tsinghua National Laboratory of Information Science and Technology,  Beijing 100084, China}

\author{Julen S. Pedernales}
\affiliation{Department of Physical Chemistry, University of the Basque Country UPV/EHU, Apartado 644, 48080 Bilbao, Spain}

\author{Lucas Lamata}
\affiliation{Department of Physical Chemistry, University of the Basque Country UPV/EHU, Apartado 644, 48080 Bilbao, Spain}

\author{Enrique Solano}
\affiliation{Department of Physical Chemistry, University of the Basque Country UPV/EHU, Apartado 644, 48080 Bilbao, Spain}
\affiliation{IKERBASQUE, Basque Foundation for Science, Maria Diaz de Haro 3, 48013 Bilbao, Spain}

\author{Gui-Lu Long}
\email[Correspondence and requests for materials should be addressed to G.L.L.: ]{gllong@tsinghua.edu.cn}
\affiliation{State Key Laboratory of Low-dimensional Quantum Physics and Department of Physics, Tsinghua University, Beijing 100084, China}
\affiliation{Tsinghua National Laboratory of Information Science and Technology,  Beijing 100084, China}
\affiliation{The Innovative Center of Quantum Matter, Beijing 100084, China}

\begin{abstract}
We measure multi-time correlation functions of a set of Pauli operators on a two-level system, which can be used to retrieve its associated linear response functions. The two-level system is an effective spin constructed from the nuclear spins of $^{1}$H atoms in a solution of $^{13}$C-labeled chloroform. Response functions characterize the linear response of the system to a family of perturbations, allowing us to compute physical quantities such as the magnetic susceptibility of the effective spin. We use techniques exported from quantum information to measure time correlations on the two-level system. This approach requires the use of an ancillary qubit encoded in the nuclear spins of the $^{13}$C atoms and a sequence of controlled operations. Moreover, we demonstrate the ability of such a quantum platform to compute time-correlation functions of arbitrary order, which relate to higher-order corrections of perturbative methods. Particularly, we show three-time correlation functions for arbitrary times, and we also measure time correlation functions at fixed times up to tenth order..
\end{abstract}

\maketitle

\section{I. Introduction}
In nature, closed quantum systems exist only as a convenient approximation. When systems are subjected to perturbations or have strong interactions with their environment, open models yield a more reliable description. A complete statistical characterization of an open quantum system unavoidably involves knowledge on the expectation value of multi-time correlations of observables, which are related to measurable quantities~\cite{Breuer02}. Time-correlation functions are at the core of optical coherence theory~\cite{Glauber63}, and can also be used for the quantum simulation of Lindbladian dynamics~\cite{DiCandia15}. A plethora of physical magnitudes, such as susceptibilities and transport coefficients, can be microscopically derived in terms of time correlation functions~\cite{Kubo57, Zwanzig65}. In a statistical approach~\cite{Kubo57}, linear response functions represent a powerful tool to compute the susceptibility of an observable to a perturbation on the system. Such functions are constructed in terms of time-correlation functions of unperturbed observables. 

Despite the ubiquity of time correlations in physics, their measurement on a quantum mechanical system is not straightforward. This difficulty lies in the fact that in quantum mechanics the measurement process disturbs the system, leaving it unreliable for a later correlated observation. Statistical descriptions typically involve an averaging of the time-correlation functions over an ensemble of particles. In such a case, it is possible to measure operator $A$ at time $t_1$ over a reduced number of particles of the ensemble, and operator $B$ at time $t_2$ over particles that were not perturbed by the first measurement. However, it is not always possible to perform measurements discriminating a subset of particles out of an ensemble. Moreover, nowadays, single quantum systems offer a high degree of controllability, which legitimates the interest in measuring time-correlation functions on single quantum systems. A solution to this puzzle can be found in algorithms for quantum computation. It is known that introducing an ancillary two-level system and performing a reduced set of controlled operations, time-correlation functions of a system can be reconstructed from single-time observables of the ancilla ~\cite{Somma02, Pedernales14, Souza11}. In this article, we measure $n$-time correlation functions for pure states, up to $n = 10$, in a highly-controllable quantum platform as is the case of nuclear magnetic resonance (NMR). Moreover, we frame these correlation functions in the context of linear response theory to compute physical magnitudes including the susceptibility of the system to perturbations. Finally, the scalability of the approach is shown to be efficient for multi-time correlation functions. 

The measurement of $n$-time correlation functions plays a significant role in the linear response theory. For instance, we can microscopically derive useful quantities such as the conductivity and the susceptibility of a system, with the knowledge of 2-time correlation functions. As an illustrative example, we study the case of a spin-$1/2$ particle in a uniform magnetic field of strength $B$ along the z-axis, which has a natural Hamiltonian $\mathcal{H}_{\rm 0}= -\gamma B \sigma_z$, where $\gamma$ is the gyromagnetic ratio of the particle. We assume now that a magnetic field with a sinusoidal time dependence $B'_0e^{-i\omega t}$ and arbitrary direction $\alpha$ perturbs the system. The Hamiltonian representation of such a situation is given by $\mathcal{H}=\mathcal{H}_{\rm 0}- \gamma B'_0 \sigma_\alpha e^{-i \omega t}$, with ${\alpha= x, y, z}$. The magnetic susceptibility of the system is the deviation of the magnetic moment from its thermal expectation value as a consequence of such a perturbation. For instance, the corrected expression for the magnetic moment in direction $\beta\  ( \mu_\beta= \gamma \sigma_\beta)$ is given by $\mu_\beta(t)= \mu_\beta(0) + \chi_{\alpha, \beta}^\omega e^{-i \omega t}$, where $\chi_{\alpha, \beta}^\omega$ is the frequency-dependent susceptibility. From linear response theory, we learn that the susceptibility can be retrieved integrating the linear response function as ${\chi_{\alpha, \beta}^\omega= \int_{-\infty}^{t} \phi_{\alpha, \beta}(t-s) e^{-i \omega (t-s)} ds}$. Moreover, the latter can be given in terms of time-correlation functions of the measured and perturbed observables, ${\phi_{\alpha, \beta}(t)=\langle [\gamma B'_0 \sigma_\alpha,  \gamma \sigma_\beta (t)] \rangle/(i \hbar)}$, where ${\sigma_\beta (t)= e^{i/\hbar H_0 t} \sigma_\beta e^{-i/\hbar H_0 t}}$, and the averaging is made over a thermal equilibrium ensemble. Notice that for a two level system, the thermal average can easily be reconstructed from the expectation values of the ground and excited states. So far, the response function can be retrieved by measuring the 2-time correlation functions of the unperturbed system $\langle \sigma_\alpha \sigma_\beta (t) \rangle $ and $\langle  \sigma_\beta (t) \sigma_\alpha\rangle$. It is noteworthy to mention that when $\alpha=\beta$, ${\langle \sigma_\alpha (t) \sigma_\alpha \rangle^*=\langle \sigma_\alpha \sigma_\alpha(t) \rangle}$, and it is enough to measure one of them. All in all, measuring two-time correlation functions from an ensemble of two level systems is not merely a computational result, but an actual measurement of the susceptibility of the system to arbitrary perturbations. Therefore, it gives us insights about the behavior of the system, and helps us characterize it. In a similar fashion, further corrections to the expectation values of the observables of the system will be given in terms of higher-order correlation functions. In this experiment, we will not only measure two-time correlation functions that will allow us to extract the susceptibility of the system, but we will also show that higher-order correlation functions can be obtained.

\section{II. The algorithm}
 
\begin{figure}[htb!]
\begin{center}
\includegraphics[width= 0.9\columnwidth]{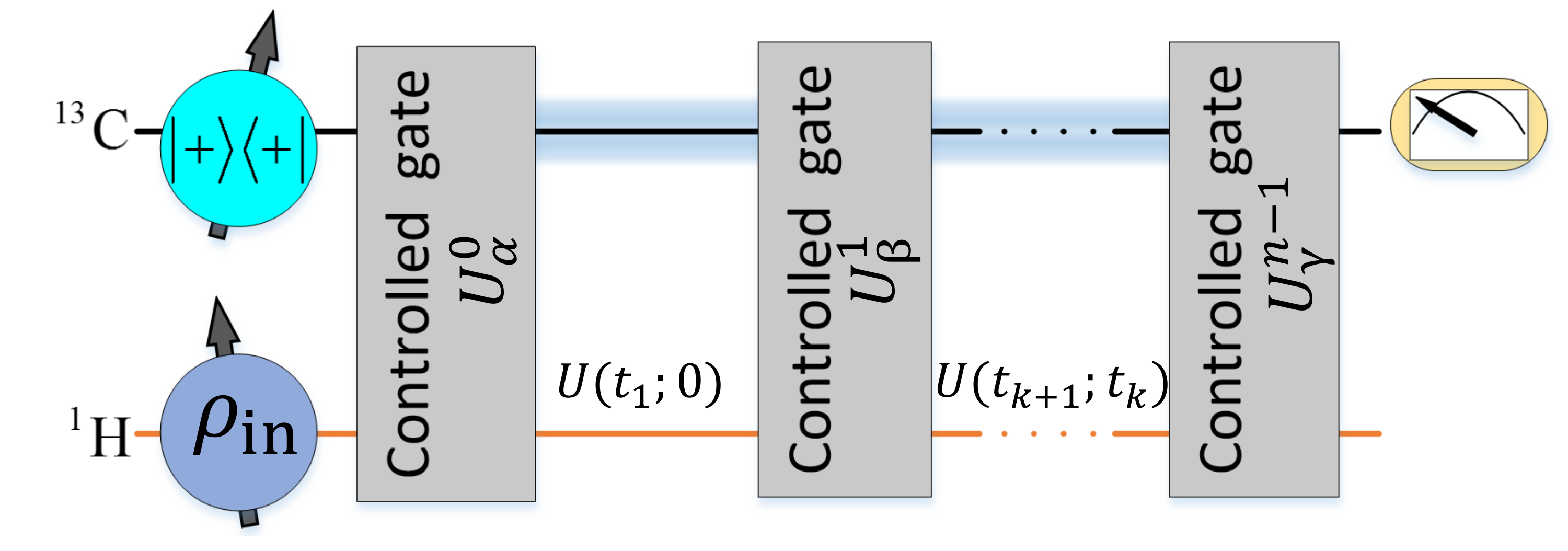}
\end{center}
\setlength{\abovecaptionskip}{-0.00cm}
\caption{\footnotesize{\textbf{Two-qubit quantum circuit for measuring general $n$-time correlation functions.} The first line is the ancilla (held by the nuclear spin of $^{13}$C), and second line is the system qubit (held by the nuclear spin of $^{1}$H). The blue zone between the different controlled gates $U^k_\alpha$ on the line of qubit A represents the decoupling of the $^{13}$C nucleus from the nuclear spin of $^{1}$H, while the latter evolves according to $U(t_{k+1};t_{k})$. The measurement of the quantities  $\langle \sigma_x \rangle$ and $\langle \sigma_y\rangle$ of the ancillary qubit at the end of the circuit will directly provide the real and imaginary values of the $n$-time correlation function for the initial state $\rho_{\rm in}=\ket{\phi}\bra{\phi}$. }}\label{circuit_total}
\end{figure}

 We will follow the algorithm introduced in Ref.~\cite{Pedernales14} to extract  $n$-time correlation functions of the form $f(t_1, ... , t_{n-1})=\bra{\phi}\sigma_\gamma (t_{n-1})...\sigma_\beta(t_1)\sigma_\alpha (0) \ket{\phi}$ from a two-level quantum system, with the  assistance of one ancillary qubit. Here, $\ket{\phi}$ is the quantum state of the two-level quantum system and $\sigma_\alpha(t)$ is a time-dependent Pauli operator in the Heisenberg picture, defined as ${\sigma_\alpha(t)=U^{\dagger}(t;0)\sigma_\alpha U(t;0)}$, where ${\alpha=x,y,z}$, and $U(t_j; t_i)$ is the evolution operator from time $t_i$ to $t_j$.  The considered algorithm of Ref.~\cite{Pedernales14} is depicted in Fig.~(\ref{circuit_total}), for the case where nuclear spins of $^{13}$C and $^{1}$H respectively encode the ancillary qubit and the two-level quantum system, and consists of the following steps: \\
$(i)$ The input state of the probe-system qubits is prepared in $\rho^{\rm CH}_{\rm in}=\ket{+}\bra{+}\otimes\rho_{\rm in}$, with $\ket{+}=(\ket{0}+\ket{1})/\sqrt{2}$ and $\rho_{\rm in}=\ket{\phi}\bra{\phi}$. \\
$(ii)$ The controlled quantum gate $U_\alpha^k=|1 \rangle \langle 1 | \otimes S_\alpha + | 0 \rangle \langle 0 | \otimes \mathbb{I}_2$ is firstly applied on the two qubits,  with $S_x=\sigma_x$, $S_y=-i\sigma_y$ and $S_z=i\sigma_z$. $\mathbb{I}_2$ is a $2 \times 2$ identity matrix. \\
$(iii)$ It follows a unitary evolution of the system qubit from $t_k$ to time $t_{k+1}$, $U(t_{k+1};t_k)$, which needs not be known to the experimenter. In our setup, we engineer this dynamics by decoupling qubit $^{13}$C and $^{1}$H, such that only the system qubit evolves under its free-energy Hamiltonian. If we were to measure the time-correlation functions for the system following a different dynamics,  the corresponding Hamiltonian should be imposed on the system at this stage of the protocol, while the system and the ancilla qubits are decoupled. Then, steps $(ii)$ and $(iii)$ will be iterated $n$ times, taking $k$ from $0$ to $n-1$ and avoiding step $(iii)$ in the last iteration. With this, all $n$ Pauli operators will be interspersed between evolution operators with the time intervals of interest $\{t_{k}, t_{k+1} \}$. \\
$(iv)$ The final state of the probe-system qubits can be written as
\begin{eqnarray}
\ket{\varphi_{\rm out}}=&& \frac{1}{\sqrt{2}} (\ket{0}\otimes U(t_{n-1};0)\ket{\phi}+\ket{1}\otimes S_\gamma U(t_{n-1};t_{n-2}) \nonumber \\
&& \cdots U(t_2;t_1)S_\beta U(t_1;0) S_\alpha \ket{\phi}).
\label{finalstate}
\end{eqnarray}
The time correlation function is then extracted as a non-diagonal operator of the ancilla, Tr($\ket{0}\langle1|\varphi_{\rm out}\rangle\bra{\varphi_{\rm out}}$). We further recall here that $| 0 \rangle \langle 1 | =( \sigma_x + i \sigma_y)/2$, such that its measurement corresponds to
\begin{equation}
{f(t_1, ..., t_{n-1})=  i^r (-i)^l (\langle \sigma_x \rangle + i \langle \sigma_y \rangle}),
\label{correlation_result}
\end{equation}
which is in general a complex magnitude, and where integers $r$ and $l$ are the occurrence numbers of Pauli operators $\sigma_y$ and $\sigma_z$ in $f(t_1, ..., t_{n-1})$. Notice that even if between the controlled operations the system can undergo dynamics that are unknown to the experimenter, the system still needs to be controllable in order for us to be able to perform the controlled-operations.

\section{III. The experiment}

\begin{figure*}[htb]
\begin{center}
\includegraphics[width= 1.6\columnwidth]{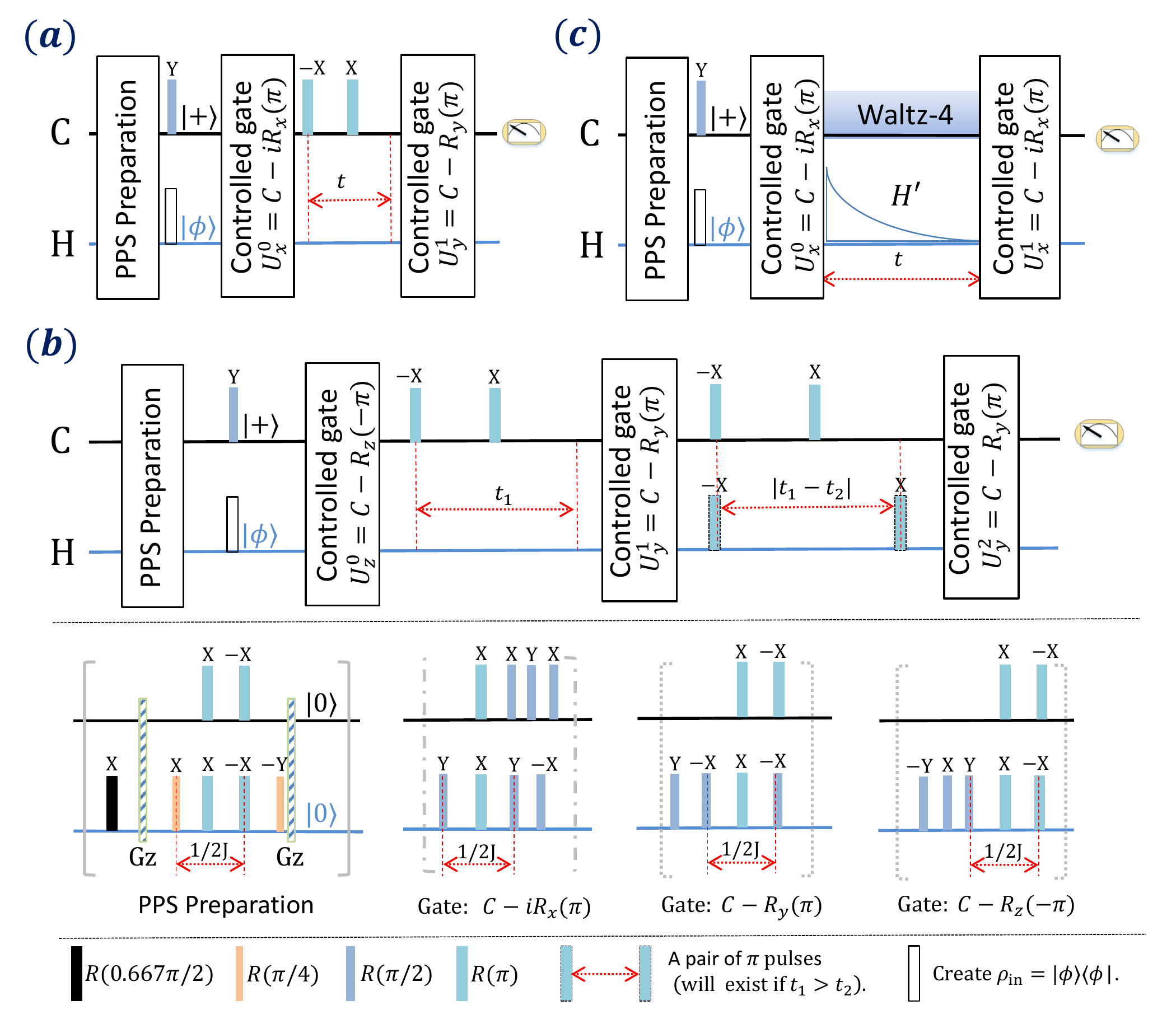}
\end{center}
\setlength{\abovecaptionskip}{-0.00cm}
\caption{\footnotesize{\textbf{NMR sequence to realize the quantum algorithm for measuring $n$-time correlation functions.} The black line and blue line mean the ancillary qubit (marked by $^{13}$C) and the system qubit (marked by $^{1}$H). All the controlled quantum gates $U^k_{\alpha}$ are decomposed into the following sequence in the bottom of the plot. Gz means a $z$-gradient pulse which is used to cancel the polarization in $x-y$ plane. (a) NMR sequence for measuring the 2-time correlation function $\langle\sigma_y(t)\sigma_x\rangle$. Other 2-time correlation functions can be similarly measured. (b) NMR sequence for measuring the 3-time correlation function $\langle\sigma_y(t_2)\sigma_y(t_1)\sigma_z\rangle$.  (c) NMR sequence for measuring the 2-time correlation function $\langle\sigma_x(t)\sigma_x\rangle$ with a time-dependent Hamiltonian $\mathcal{H'}(t)=500e^{-300t}\pi\sigma_y$. The method to decouple the interaction between $^{13}$C and $^{1}$H nuclei is Waltz-4 sequence. }}\label{circuit_detail}
\end{figure*}

\begin{figure}[htb]
\begin{center}
\includegraphics[width= 0.9\columnwidth]{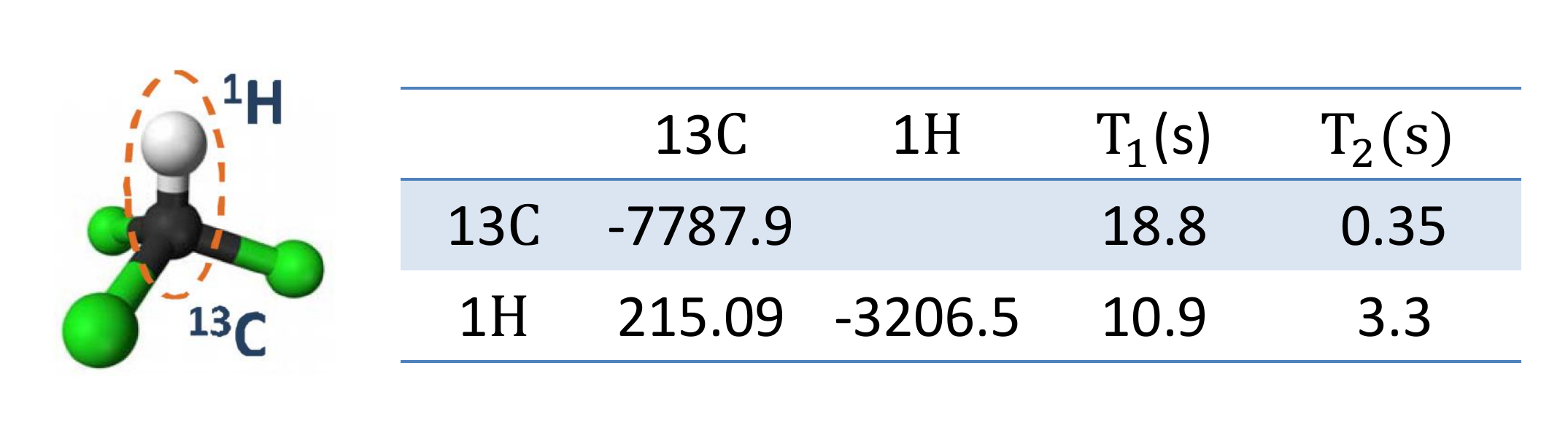}
\end{center}
\setlength{\abovecaptionskip}{-0.00cm}
\caption{\footnotesize{\textbf{Molecular structure and relevant parameters of $^{13}$C-labeled Chloroform.} Diagonal elements and off-diagonal elements in the table provide the values of the chemical shifts (Hz) and $J$-coupling constant (Hz) between $^{13}$C and $^{1}$H nuclei of the molecule. The right table also provides the longitudinal time $T_1$ and  transversal relaxation $T_2$ , which can be measured using the standard inversion recovery and Hahn echo sequences. }}\label{molecule}
\end{figure}

We will measure $n$-time correlation functions of a two-level quantum system with the  assistance of one ancillary qubit by implementing the quantum circuit shown in Fig.~(\ref{circuit_total}). Experiments are carried out using NMR \cite{Cory00,Havel02,Suter08}, where the sample used is $^{13}$C-labeled chloroform. Nuclear spins of $^{13}$C and $^{1}$H encode the ancillary qubit and the two-level quantum system, respectively. With the weak coupling approximation, the internal Hamiltonian of $^{13}$C-labeled chloroform is
\begin{equation}
\mathcal{H}_{\rm int}=-\pi (\nu _1-\nu^o _1)\sigma_z^1-\pi (\nu _2-\nu^o _2)\sigma_z^2+\frac{1}{2} \pi J_{12} \sigma_z^1 \sigma_z^2,
\label{hamiltonian}
\end{equation}
where $\nu_j$ ($j=1,2$) is the chemical shift, and $\emph{J}_{12}$ is the $J$-coupling strength. Fig. \ref{molecule} shows the molecular structure and properties of the sample. While $\nu^o _1$ and $\nu^o _2$ are reference frequencies of $^{13}$C and $^{1}$H, respectively.  We set $\nu _1=\nu^o _1$ and $\nu _2-\nu^o _2=\bigtriangleup\nu$ such that the natural Hamiltonian of the system qubit is $\mathcal{H}_{0}=-\pi \bigtriangleup \nu \sigma_z$. The detuning frequency $\bigtriangleup\omega$ is chosen as hundreds of Hz to assure the selective excitation of different nuclei via hard pulses. All experiments are carried out on a Bruker AVANCE 400MHz spectrometer at room temperature. 

\begin{figure}[tbp]
\begin{center}
\includegraphics[width= 0.9\columnwidth]{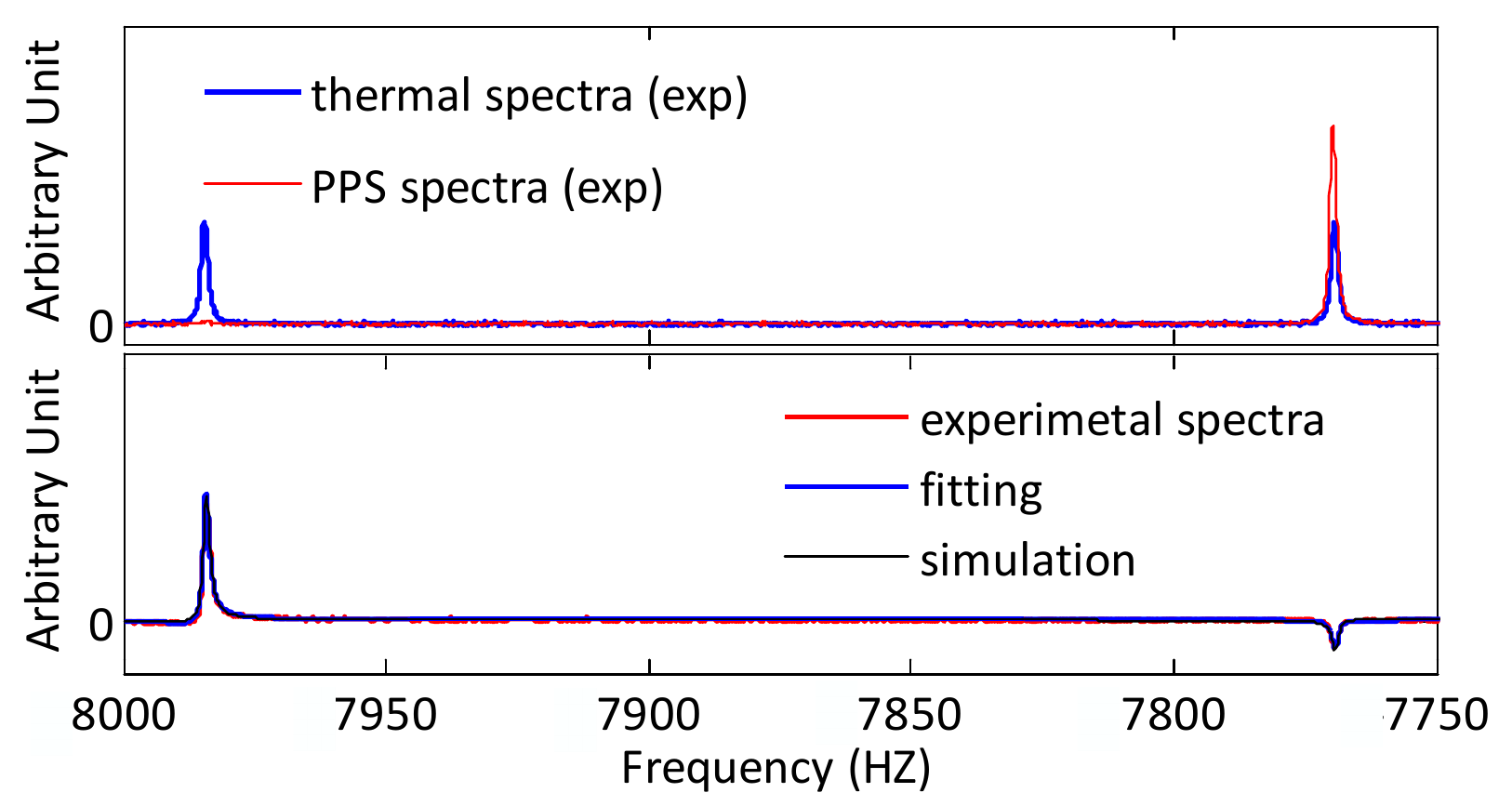}
\end{center}
\setlength{\abovecaptionskip}{-0.00cm}
\caption{\footnotesize{\textbf{Experimental spectra of $^{13}$C nuclei.} The blue line of the top plot shows the observed spectrum after a $\pi/2$ pulse is applied on $^{13}$C nuclei in the thermal equilibrium state. The signal measured after applying a $\pi/2$ pulse following the preparation of the PPS is shown by the red line of the top plot. The bottom plot shows the spectrum when we measure $\mathcal{M}^n_{xy}$ ($n=2$). The red, blue and black lines represent the experimental spectra, fitting results and corresponding simulations, respectively.}}
\label{spectra}
\end{figure}

It is widely known that the thermal equilibrium state of a two-spin CH NMR ensemble is a highly-mixed state with the following structure
\begin{equation}
\mathcal{\rho}_{eq}\approx \frac{1-\epsilon}{4}\mathbb{I}_4+\epsilon(\frac{1}{4}\mathbb{I}_4+\sigma^1_z+4\sigma^2_z ).
\end{equation}
Here, $\mathbb{I}_4$ is a $4 \times 4$ identity matrix and $\epsilon\approx10^{-5}$ is the polarization at room temperature. Given that $\mathbb{I}_4$ remains unchanged and that it does not contribute to the NMR spectra, we consider the deviation density matrix $\bigtriangleup \rho=0.25\mathbb{I}_4+\sigma^1_z+4\sigma^2_z $ as the effective density matrix describing the system. The deviation density matrix can be initialized in the pure state $\ket{00}\bra{00}$ by the spatial averaging technique \cite{Cory97,Cory98,Knill98}, transforming the system into the so-called pseudo-pure state (PPS). The top plot in Fig. \ref{spectra} shows the spectra of the thermal equilibrium and pseudo pure states. Arbitrary input states $\rho^{\rm CH}_{\rm in}$ can be easily created by applying local single-qubit rotation pulses after the preparation of the PPS.

In Step $(ii)$, all controlled quantum gates $U^k_\alpha$ are chosen from the set of gates $\{C-R^2_z(-\pi), C-iR^2_x(\pi), C-R^2_y(\pi) \}$ \cite{Xin15,Vandersypen05,Bendall83}. The notation $C-U$ means operator $U$ will be applied on the system qubit only if the ancilla qubit is in state $\ket{1}\bra{1}$, while $R^j_{\hat{n}}(\theta)$ represents a single-qubit rotation on qubit $j$ along the $\hat{n}$-axis, with the rotation angle $\theta$.  We decompose the family of controlled quantum gates $U^k_\alpha$ = $\{C-R^2_z(-\pi), C-iR^2_x(\pi), C-R^2_y(\pi) \}$ in the following way
\begin{eqnarray}
&& C-R^2_z(-\pi) = U(\frac{1}{2J})R^2_z(-\frac{\pi}{2}) , \nonumber \\
&& C-iR^2_x(\pi) = \sqrt{i}R^1_z(\frac{\pi}{2})R^2_z(-\frac{\pi}{2})R^2_x(\frac{\pi}{2})U(\frac{1}{2J})R^2_y(\frac{\pi}{2}) , \nonumber \\
&& C-R^2_y(\pi) = R^2_x(\frac{\pi}{2})U(\frac{1}{2J})R^2_x(-\frac{\pi}{2})R^2_y(\frac{\pi}{2}).
\label{decompose}
\end{eqnarray}
Here, $U(\frac{1}{2J})$ is the $J$-coupling evolution $e^{-i\pi\sigma^1_z\sigma^2_z/4}$. Moreover, any $z$-rotation $R_z(\theta)$ can be decomposed in terms of rotations around the $x$ and $y$ axes, $R_z(\theta)=R_y(\pi/2)R_x(-\theta)R_y(-\pi/2)$. It is worth mentioning that these decompositions are in terms of propagators and not pulses, and therefore that these propagators should be applied from right to left.

We will now apply the described algorithm to a collection of situations of physical interest. These include two-time correlation functions of a system evolving under time-independent and time-dependent Hamiltonians, as well as three-time correlation functions. In Fig.~\ref{circuit_detail} we give the detailed NMR sequences employed for the measurement of the time-correlation functions in each of these cases. More especifically, Fig. \ref{circuit_detail} (a) shows the experimental sequence for measuring $\langle\sigma_y(t)\sigma_x\rangle$. Other time-correlation functions, like  $\langle\sigma_x(t)\sigma_y\rangle$, $\langle\sigma_y(t)\sigma_y\rangle$ or  $\langle\sigma_x(t)\sigma_z\rangle$, can be measured in a similar fashion by replacing the corresponding controlled quantum gates. Figure \ref{circuit_detail} (b) describes the NMR sequence for measuring the three-time correlation function $\langle\sigma_y(t_2)\sigma_y(t_1)\sigma_z\rangle$ for different values of $t_1$ and $t_2$. Here we used a pair of $\pi$ pulses, which change the sign of the Hamiltonian $\mathcal{H}_0$, if $t_1$ is greater than $t_2$. Finally, Fig. \ref{circuit_detail} (c) illustrates the NMR sequence corresponding to the measurement of the time-correlation function $\langle\sigma_x (t) \sigma_x\rangle$ with a time-dependent Hamiltonian of the form $\mathcal{H'}(t)=500e^{-300t}\pi\sigma_y$. The dynamics corresponding to this Hamiltonian are generated by a time-dependent radio-frequency pulse applied on the resonance of the nuclear spin of $^1$H, that is to say on the system qubit.

\begin{figure*}[htb]
\begin{center}
\includegraphics[width=1.4 \columnwidth]{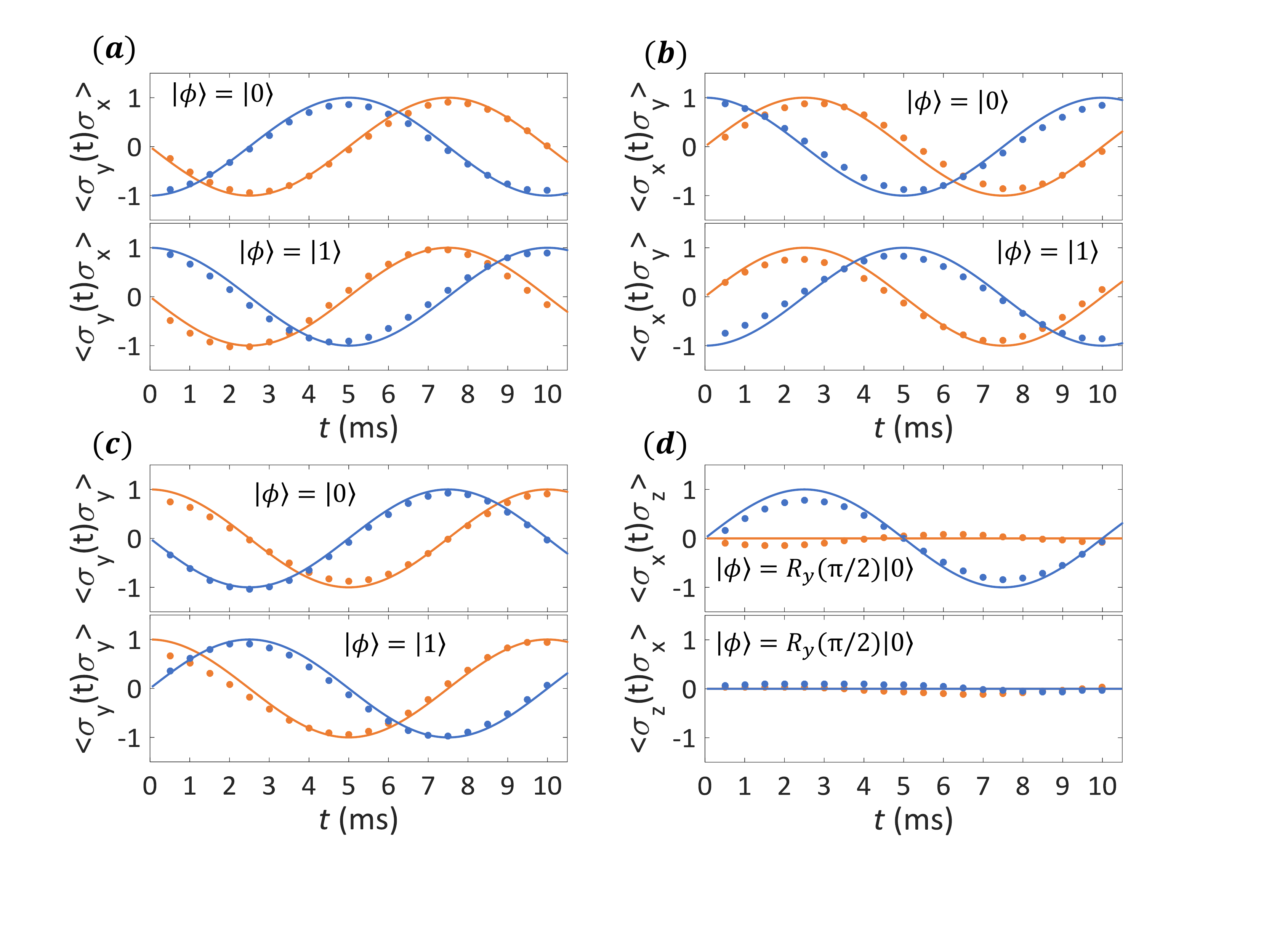}
\end{center}
\setlength{\abovecaptionskip}{-0.00cm}
\caption{\footnotesize{\textbf{Experimental results (dots) for 2-time correlation functions.} In this case, only two controlled quantum gates $U^0_\alpha$ and $U^1_\beta$ are applied  with an interval of $t$. For example, $U^0_\alpha$ and $U^1_\beta$ should be chosen as $C-iR^2_x(\pi)$ and $C-R^2_y(\pi)$, respectively, to measure the 2-time correlation function $\langle\sigma_y(t)\sigma_x\rangle$. $t$ is swept from $0.5$ms to $10$ms with a $0.5$ms increment. The input state of $^{1}$H nuclei $\rho_{\rm in}=\ket{\phi}\bra{\phi}$ is shown on each diagram. All experimental results are directly obtained from measurements of the expectation values of  $\langle\sigma_x\rangle$ and $\langle\sigma_y\rangle$ of the ancillary qubit. The orange and blue results respectively mean the real and imagine part of the observed 2-time correlation functions.}}\label{classA}
\end{figure*}

In Fig.~(\ref{classA}), we show the measured two-time correlation functions $\langle\sigma_\alpha(t)\sigma_\beta(0) \rangle$ for a collection of $\alpha$ and $\beta$, and different initial states. In this experiment the two-level system was evolving under the Hamiltonian $\mathcal{H}_{0}=-100\pi\sigma_z$.  The observed oscillations correspond to the rotation of the two-level system along the z-axis of its Bloch sphere, as dictated by the evolution Hamiltonian. Consistently, the bottom plot of Fig.~\ref{classA} (d) shows no oscillations, as the time-dependent operator in this case is $\sigma_z(t)$, which is aligned with the oscillation axis. The plotted times correspond to the time-scales of the implemented dynamics. The chosen millisecond time-range is especially convenient, as the decoherence effects become significant only at longer times. In the experiment, $\mathcal{H}_{0}$ is realized by setting $\nu _1=\nu^o_1$ and $\nu _2-\nu^o_2=100$ Hz in Eq.~(\ref{hamiltonian}). A rotation pulse $R^1_y(\pi/2)$ is applied on the first qubit after the PPS preparation to create $\rho^{\rm CH}_{\rm in}=\ket{+}\bra{+}\otimes\ket{0}\bra{0}$. Similarly, a $\pi$ rotation on the second qubit is additionally needed to prepare $\rho^{\rm CH}_{\rm in}=\ket{+}\bra{+}\otimes\ket{1}\bra{1}$ as the input state of the ancilla-system compound, or alternatively a $R_y^2(\pi/2)$ rotation to generate the initial state $\rho_{\rm in}^{\rm CH}= | + \rangle \langle + | \otimes | + \rangle \langle + |$.

These correlation functions are enough to retrieve the response function for a number of physical situations corresponding to different magnetic moments and applied fields. On the other hand, extracting correlation functions for initial states $| 0 \rangle$ and $| 1 \rangle $ will allow us to reconstruct such correlation functions for a thermal state of arbitrary temperature.

\begin{figure}[htb]
\begin{center}
\includegraphics[width= 0.9\columnwidth]{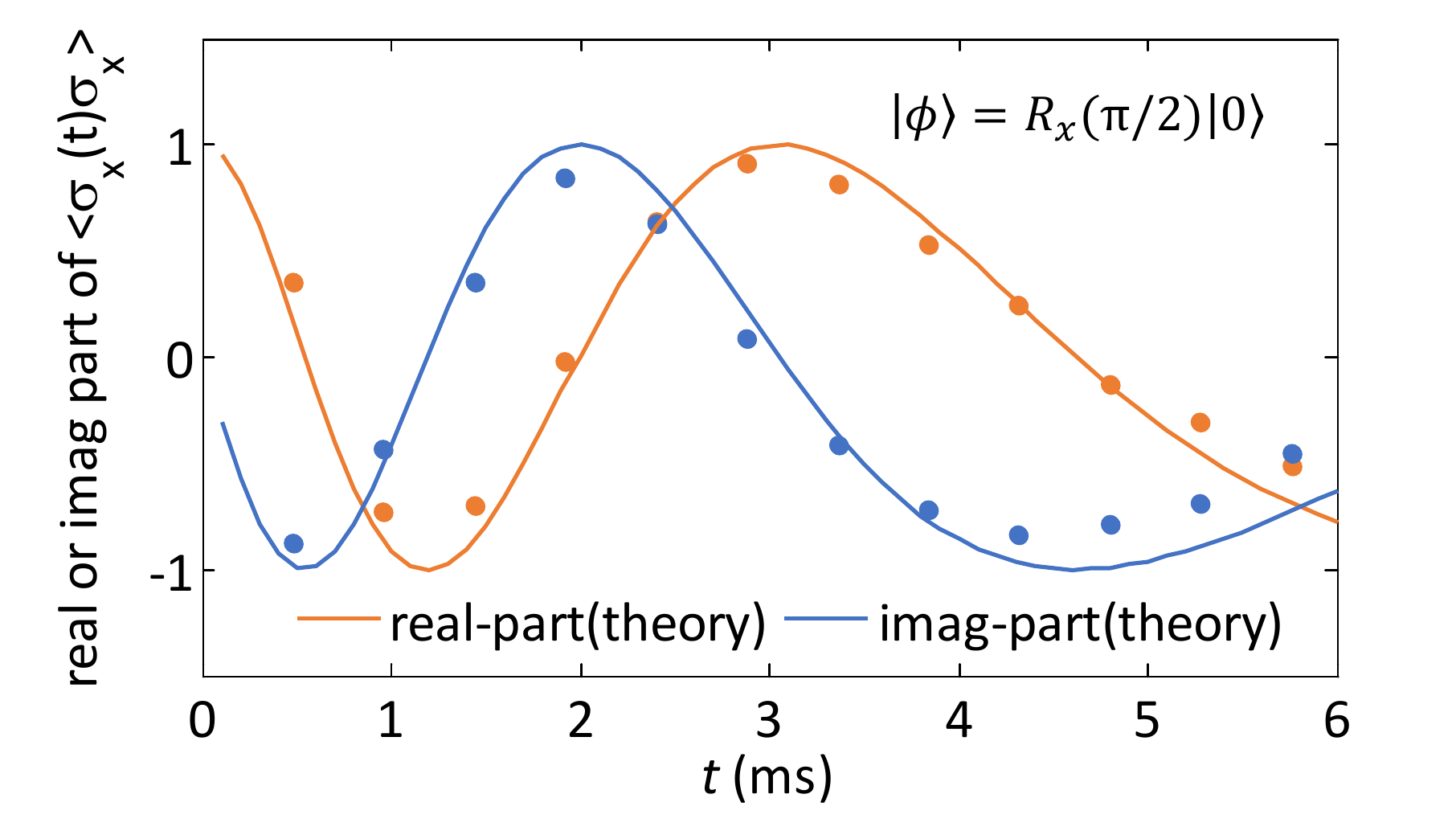}
\end{center}
\setlength{\abovecaptionskip}{-0.00cm}
\caption{\footnotesize{\textbf{Experimental results (dots) for a 2-time correlation function of the $^1$H nuclei evolving under a time-dependent Hamiltonian.} For this experiment, the $^1$H nuclei have a natural Hamiltonian $\mathcal{H}_{0}=0$ and an initial state $\ket{\phi}=R_x(\pi/2)\ket{0}$. An evolution $U(t;0)$ between $U^0_x$ and $U^1_x$ is applied on the system, which is described by the evolution operator $e^{-i\int_{0}^t \mathcal{H'}(s) ds}$ with $\mathcal{H'}(s)=500e^{-300s}\pi\sigma_y$. $t$ is changed from $0.48$ms to $5.76$ms with a $0.48$ms increment per step.}}\label{classD}
\end{figure}

In Fig.~\ref{classD} we show the measured time-correlation fuction  $\langle\sigma_x (t) \sigma_x\rangle$ for the initial state $(| 0 \rangle-i| 1 \rangle)/\sqrt{2}$ evolving under the time dependent Hamiltonian $\mathcal{H'}(t)=500e^{-300t}\pi\sigma_y$. For this, we set $\nu _1=\nu^o_1$ and $\nu _2=\nu^o_2$ in Hamiltonian $\mathcal{H}_{int}$ in Eq.~(\ref{hamiltonian}), making the system free Hamiltonian $\mathcal{H}_{0}=0$. The initial state $\ket{\phi}=R_x(\pi/2)\ket{0}$ can be prepared by using a rotation pulse $R_x(\pi/2)$ on the initial PPS. Two controlled quantum gates $U^0_x=C-iR^2_x(\pi)$ and $U^1_x=C-iR^2_x(\pi)$ are applied  with a time interval $t$. A decoupling sequence Waltz-4~\cite{Widmaier98, Shaka83a, Shaka83b} is used to cancel the interaction between the $^{13}$C and $^{1}$H nuclei during the evolution between the controlled operations. During the decoupling period, a time-dependent radio-frequency pulse is applied on the resonance of the system qubit $^1$H to create the Hamiltonian, as explained above. From a physical point of view, this kind of correlations would be descriptive of a situation where the system is in a magnetic field with an intensity that is decaying exponentially in time, that is, the unperturbed system Hamiltonian turns now into a time-dependent $\mathcal{H'}= \gamma B_0 e^{-a t} \sigma_y$. A degradation in agreement between experiment and theory is observed at the upper end of times in Fig.~4. This is due to the cumulative effects of decoherence mechanisms and the power attenuation of the employed radio-frequency pulses at long times, which results in a weak NMR response of the nuclei and as a consequence in more imprecise spectroscopic results.

\section{IV. Third time correlations}

When the perturbation is not weak enough, for instance when the radiation field applied to a material is of high intensity, the response of the system might not be linear. In such situations,  higher-order response functions, which depend in higher-order time-correlation functions, will be needed to account for the non-linear corrections~\cite{Kubo57, Peterson67}. For example, the second order correction to an observable $B$ when the system suffers a perturbation of the type $H(t)= H_0 + A F(t)$ would be given by ${\Delta B^{(2)}= \int_{-\infty}^t \int_{-\infty}^{t_1} \langle [B(t), [A(t_1), A(t_2)] \rangle F(t_1) F(t_2) dt_1 dt_2 }$. 

\begin{figure}[htb]
\begin{center}
\includegraphics[width= \columnwidth]{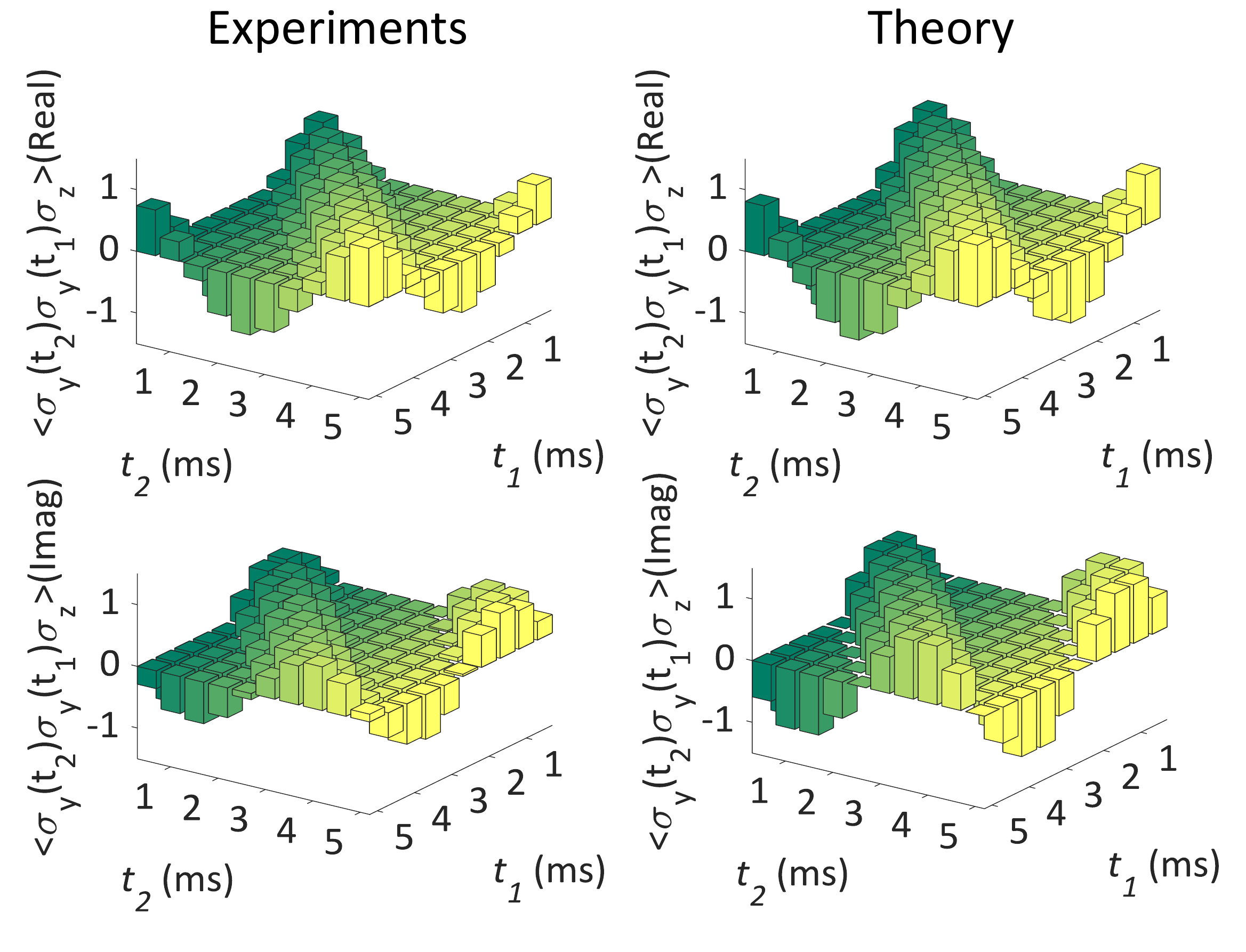}
\end{center}
\setlength{\abovecaptionskip}{-0.00cm}
\caption{\footnotesize{\textbf{Experimental results for the $3$-time correlation functions.} We plot $\mathcal{M}^3_{zyy} = \langle\sigma_y(t_2)\sigma_y(t_1)\sigma_z\rangle$ for $t_1$ and $t_2$ going from 0.5ms to 5ms with 0.5ms time step, showing the agreement of experimental results with theoretical predictions. The quantum circuit for measuring $\mathcal{M}^3_{zyy}$ includes three controlled quantum gates $U^0_z$,  $U^1_y$ and $U^2_y$, which are experimentally implemented by hard pulses. }}\label{classB}
\end{figure}

In Fig.~(\ref{classB}), we show real and imaginary parts of 3-time correlation functions as compared to their theoretically expected values. We measure the 3-time correlation function $\langle\sigma_y(t_2)\sigma_y(t_1)\sigma_z\rangle$ versus $t_1$ and $t_2$.  Like in the case of the two-time correlation functions, the oscillatory behavior of the measured three-time correlation functions reflects the rotation of the two-level system along the z-axis of its Bloch sphere. In this case, we simulate the system-qubit free Hamiltonian $\mathcal{H}_{0}=-200\pi\sigma_z$ for the initial state $\rho_{\rm in}=\ket{0}\bra{0}$. For this, we set $\nu _1=\nu^o_1$ and $\nu _2-\nu^o_2=200$ Hz in Eq.~(\ref{hamiltonian}). The $J$-coupling term of Eq.~(\ref{hamiltonian}) will be canceled by using a refocusing pulse in the circuit.  Three controlled quantum gates $U^0_\alpha$,  $U^1_\beta$ and $U^2_\gamma$ should be chosen as $C-R^2_z(-\pi)$, $C-R^2_y(\pi)$ and $C-R^2_y(\pi)$. The free evolution of the $^{1}$H nuclei between $U^0_\alpha$ and $U^1_\beta$ is given by the evolution operator $e^{-i\mathcal{H}_{0}t_1}$. Accordingly,  the free evolution of the $^{1}$H nuclei between $U^1_\beta$ and $U^2_\gamma$ is given by $e^{-i\mathcal{H}_{0}(t_2-t_1)}$. However, when $t_2$<$t_1$, we perform the evolution $e^{-i(-\mathcal{H}_{0})(t_1-t_2)}$ by inverting the phase of the Hamiltonian $\mathcal{H}_{0}$, which is realized by using a pair of  $\pi$ pulses at the beginning and at the end of the evolution~\cite{Vandersypen05}. \\

For testing scalability, we also measure higher-order time-correlation functions, up to $n=10$. In this case, we consider the free Hamiltonian ${\mathcal{H}_{0}=-100\pi\sigma_z}$ and the input state $\ket{\phi}=R_x(1.41\pi/2)\ket{0}$, and we measure the following high-order correlation functions as a function of the correlation order $n$, with time intervals $t_{n-1}=0.3(n-1)$ ms,
\begin{equation}
\begin{array}{l}
\mathcal{M}^n_{xx}=\langle\sigma_x(t_{n-1})\sigma_x(t_{n-2})...\sigma_x(t_{1})\sigma_x\rangle,\\
\mathcal{M}^n_{xy}=\langle ... \sigma_x(t_{2m})\sigma_y(t_{2m-1})...\sigma_y(t_{1})\sigma_x\rangle.
\end{array}
\end{equation}

Here, the superscripts and subscripts of $\mathcal{M}$ are the order of the correlation and the involved Pauli operators, respectively. Index $m$ runs from $1$ to $(n-1)/2$ for odd $n$, and to $n/2$ for even $n$. The quantum circuit used to measure $\mathcal{M}^n_{xx}$ and $\mathcal{M}^n_{xy}$ is based in the gradient ascent pulse engineering (GRAPE) technique \cite{Khaneja05,Ryan08}, which is designed to be robust to the static field distributions ($T_2^{*}$ process) and RF inhomogeneities. 

\begin{figure}[htb]
\begin{center}
\includegraphics[width= \columnwidth]{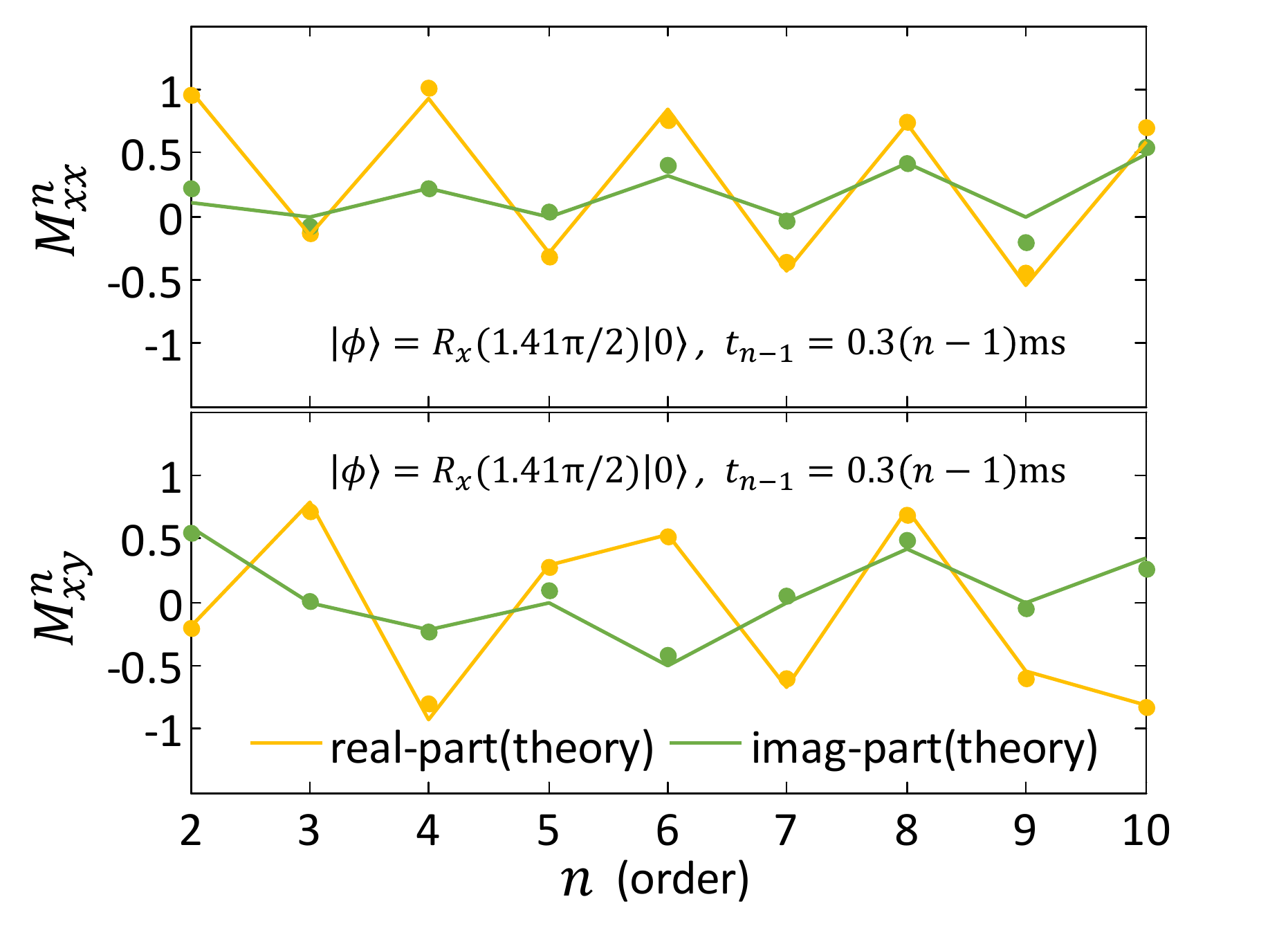}
\end{center}
\setlength{\abovecaptionskip}{-0.00cm}
\caption{\footnotesize{\textbf{Experimental results (dots) for high-order time-correlation functions $\mathcal{M}^n_{xx}$ and $\mathcal{M}^n_{xy}$ ($n=2,3,...,10$).} $n$ controlled quantum gates $U^0_\alpha$,$U^1_\beta$,...,$U^{n-1}_\gamma$ are sequentially applied  with a time interval $\bigtriangleup t=0.3$ms. Refocusing pulses are used to decouple the interaction between the nuclei of $^{13}$C and $^{1}$H, during the time intervals $\bigtriangleup t$ between gates. For this experiment, we use the GRAPE pulsed technique to implement the quantum circuit. Using hard pulses like in the previous experiments would result in poor quality of the measured data due to the high number of pulses required and the cumulative effect of their imperfections.}}\label{classC}
\end{figure}

In Fig. \ref{classC}, we show the measured results for high-order time-correlation functions, demonstrating the scalability of the technique and its high accuracy even for a $10$-time correlation function. With this we demonstrate that high-order correlation functions are efficiently accessible in NMR via our algorithm. In general, high-order time-correlation functions correspond to high-order corrections in perturbation theories. In our example, the jagged pattern of the measured data with the order of the time-correlation function can be explained in terms of each order corresponding to measurements in different axes of the Bloch sphere.

For systems of bigger size and complex dynamics our technique should be equally valid, and would be useful for computational purposes, when the dynamics of the system is not reproducible by classical means. In this case, the algorithm would also work with a single ancillary qubit, however, the controlled gates would pass from two-qubit gates to multi-qubit gates, which have been little studied in NMR. Nevertheless, multi-qubit gates like the M\o lmer-S\o rensen gate can always be efficiently decomposed into a circuit of c-NOT gates~\cite{Pedernales14}, allowing for the measurement of multi-qubit time correlations in NMR. Meanwhile, the bottom plot in Fig. \ref{spectra} shows NMR spectra which is created after we measure the 2-time correlation function $\mathcal{M}^n_{xy}$ ($n=2$). 
 
For all cases here discussed, experimental data shows a high degree of agreement with the theoretical predictions. Error bars are not shown, as they are always smaller than the used dots themselves. The dephasing times $T_2$ of our spin-qubits are of the order of seconds, while the experimental time of a whole sequence is at most of $10$~ms, allowing us to ignore the effect of dephasing effects during the experiment. In our setup, the sources of errors are related to the initialization of the PPS, data-fitting, and imperfections in the width of the employed hard pulses. Moreover, the latter effect is cumulative and can result in a snowball effect. Additionally, factors such as RF inhomogeneities, bring in a signal loss.

\section{Conclusion}

We have shown that the measurement of time-correlation functions of arbitrary order in NMR is an efficient task, and that it can be used to obtain the linear response function of the system. Although the linear response function could be calculated indirectly with a precise determination of the Hamiltonian parameters of the system, this experiment can be considered its first direct measurement in NMR. For systems of bigger size and complex dynamics, the indirect estimation of this magnitude would become intractable, as an analytical or numerical solution of the dynamics is always required. However, a direct detection would still be possible following the ideas demonstrated in this experiment. Not only that, in this work, we have demonstrated that such magnitudes can be experimentally retrieved with high accuracy. This will be of interest for physicist and engineers, either to characterize systems that follow computationally intractable dynamics, or to use them for computation purposes, opening the door to the quantum simulation of physical models where time correlations play a central role. It is generally accepted, that NMR platforms scale poorly, and there is no indication that this will change in the foreseeable future. However, the central ideas demonstrated in this experiment do not rely on any property which is exclusive of NMR platforms. Therefore, it is our believe that other more scalable quantum platforms may extend the protocol demonstrated here to systems of arbitrary size, where a single ancillary qubit will always suffice.

\section{V. Acknowledgments}
T. X. and G.-L. L. are grateful to the following funding sources: National Natural Science Foundation of China under Grants No. 11175094 and No. 91221205; National Basic Research Program of China under Grant No. 2015CB921002. J. S. P., L. L. and E. S. acknowledge support from Spanish MINECO FIS2015-69983-P; UPV/EHU UFI 11/55; Ram\'on y Cajal Grant RYC-2012-11391; Basque Government Grant IT986-16; UPV/EHU Project EHUA14/04 and a UPV PhD grant.

\end{document}